\documentstyle[preprint,epsfig,eqsecnum,aps]{revtex}
\tighten

\newcommand{\beq}{\begin{equation}}
\newcommand{\eeq}{\end{equation}}
\newcommand{\beqn}{\begin{eqnarray}}
\newcommand{\eeqn}{\end{eqnarray}}
\newcommand{\beqa}{\begin{eqnarray}}
\newcommand{\eeqa}{\end{eqnarray}}
\newcommand{\bi}{\bibitem}
\newcommand{\n}{\newline}
\newcommand{\nn}{\nonumber\\}
\newcommand{\bc}{\begin{center}}
\newcommand{\ec}{\end{center}}

\begin{document}
\preprint{LPTENS-95/45}
\title{Phase space geometry and slow dynamics}
\author{Jorge Kurchan and Laurent Laloux}
\address{ 
CNRS-Laboratoire de Physique Th\'{e}orique de l'Ecole Normale 
Sup\'{e}rieure$^1$  \\
24, rue Lhomond; 75231 Paris Cedex 05; France.\\
e-mail: kurchan@physique.ens.fr, laloux@physique.ens.fr \\[1cm]
{\footnotesize $^1$  Unit\'{e} propre du CNRS (UP 701) associ\'{e}e
\`{a} l'ENS  et \`{a} l'Universit\'{e} Paris-Sud.} 
} 
\date{\today}

\maketitle

\begin{abstract}
We describe a non-Arrhenius mechanism for slowing down of dynamics that is
inherent to the high dimensionality of the phase space.
We show that such a  mechanism is at work both in a family of mean-field
spin-glass models without any domain structure and in the case of
ferromagnetic domain growth.\\
The marginality of spin-glass dynamics, as well as the existence of a
`quasi equilibrium regime' can be understood within this scenario. \\
We discuss the question of ergodicity in an out-of equilibrium situation.
\end{abstract}
\narrowtext

\section{Introduction}
\renewcommand{\theequation}{\thesection.\arabic{equation}}
\label{sect0}

Many systems of physical interest are  out of equilibrium throughout the 
observation times after preparation.
The fact that  a system rather than reaching the Gibbs-Boltzmann 
equilibrium measure remains in a regime of slow dynamics can be attributed 
to various causes.
A clear example is the case in which there are   domains of  different
ordered phases growing at the expense of each other, as when a ferromagnet 
is quenched to the low temperature phase. 
Another rather different scenario is when the phase-space  has traps of long
lifetimes, which the system leaves without visiting again.

Spin-glasses (and also structural glasses) are known  to have properties
that depend on the `age' after the quench \cite{St,Aging},
and hence the possibility that they are in equilibrium is ruled out.
Several explanations have been proposed to account for their slow dynamics,
based on domain growth ideas \cite{Fihu}, on a phase space with
traps \cite{Bo} and on a percolation-like picture in phase-space 
\cite{Siho,Nest}.
The latter two scenarios are  low-dimensional in the sense
that they work equally well in a low dimensional (though infinite) phase-space.

The purpose of this paper is to argue, with some examples,
that just  as equilibrium thermodynamical properties 
such as the existence of macroscopic non-fluctuating quantities are 
a direct consequence of the infinite dimensionality of phase-space 
(irrespective of the physical dimensionality $D$), there are also in the 
out of equilibrium dynamics aspects that are inherent to the geometry  
of infinite-dimensional (phase) spaces. 

We shall first describe these rather generic geometric features, and then
show explicitly how they lead to slow dynamics, even in the absence
of metastable states. We shall see that they apply to both ferromagnetic 
domain  growth and to a family of mean-field spin-glass models which does not 
have any domain structure.
In both cases we shall concentrate on `long but finite times':
the limit $N \rightarrow \infty $ (or $V \rightarrow \infty$)
is made before the limit $t \rightarrow \infty$.

In order to have a well defined landscape in which a deterministic
dynamics takes place, we shall restrict our discussion to zero or
near-zero temperatures.  We believe that the mechanism we shall
describe is also at work in the  case of higher temperatures, 
possibly cooperating with other specifically non-zero temperature
mechanisms such as barrier crossing.  The three models we shall use as
examples have the property that their dynamics at zero and at low
temperature are essentially the same.

The problem with extending our geometrical discussion to finite temperatures
is that at present we do not know exactly the geometry of {\em what} we should 
look into (that is, short of the whole Hilbert space of the 
Fokker-Planck equation).
In this respect, a usual thing is to  have in mind a {\em free}-energy 
landscape in terms of variables  representing the evolution of a 
probability packet. Whatever the procedure for the construction
of such a landscape, the  implicit assumption is that the dynamics is 
{\em deterministic} in these variables (otherwise the original energy 
landscape would be as good).

There seems to be, however, 
discouraging evidence for this approach, at least for glassy systems:
It has been shown \cite{Babume} that a set of trajectories 
that are forced to coincide up to any given finite time, 
and are then subjected to different thermal noises will eventually
diverge to distant places of the phase space, while with a deterministic 
approach one would conclude that the they evolve together. 
In other words, a probability packet that is out of equilibrium
is destroyed by the evolution.

We shall concentrate on systems with a smooth  energy-density landscape 
with no relevant infinite energy density  configurations. 
Let us define the normalized square phase-space distance between two 
configurations $s_i^a$, $s_i^b$:
\beq
B(a,b)= \frac{1}{N} \sum_{i=1}^N (s_i^a - s_i^b)^2
\label{B1}
\eeq
or, for two fields $\phi^a(x)$, $\phi^b(x)$:
\beq
B(a,b)= \frac{1}{V} \int_0^L d^D x (\phi^a(x)-\phi^b(x))^2
\label{B2}
\eeq
The correlation function is introduced in the usual way,
\beq
B(a,b)=C(a,a)+C(b,b)-2C(a,b)
\eeq
We shall say that a system has well-separated  energy minima if
$B(a,b)$ between  any two minima is 
 an $O(1)$ quantity, or:
\beq
\frac{C(a,b)}{[C(a,a)C(b,b)]^{1/2}}<1
\label{sepa}
\eeq
In the case of non-zero temperature the corresponding question is whether
the correlation between magnetizations in two states is smaller than  one:
\beq
\frac{\sum_i m^a_i m^b_i}{ \sqrt{\sum_i (m^a_i)^2 } \sqrt{\sum_i (m_i^b)^2}} < 1
\label{separatio}
\eeq

Some examples of systems with well-separated minima are the ferromagnet
and ferromagnetic Potts models in any dimension and mean-field
spin-glasses  with  finitely many breakings. Instead, mean-field
spin-glasses with infinitely many levels of replica symmetry breaking
do not satisfy this condition.  Our discussion is mainly directed at
systems of the first kind.

We shall exemplify the geometrical properties we discuss here with three models.
The first one is a ferromagnetic domain-growth problem 
(see \cite{Bray} for a review). The energy is of the Landau type
\beq
E(\phi)= \int d^D x \; \left( \frac{1}{2} (\nabla \phi )^2 + V(\phi) \right)
\label{Landau}
\eeq
where $V$ has a double-well structure with minima at $\phi = \pm 1$ (Fig.1).
We take $V(\pm 1)=0$, and
consider a system of side $L$ and periodic boundary conditions.
In order not to have regions of phase-space with infinite energy 
densities we shall assume that there is an ultraviolet cutoff.
The dynamics is gradient descent:
\beq
\frac{\partial \phi}{\partial t} = - \frac{\delta E}{\delta \phi(x)}=
-\nabla^2 \phi(x) - V'(\phi(x))
\label{??}
\eeq
starting from a random configuration.

\begin{center}
   \parbox{8cm}{\epsfig{file=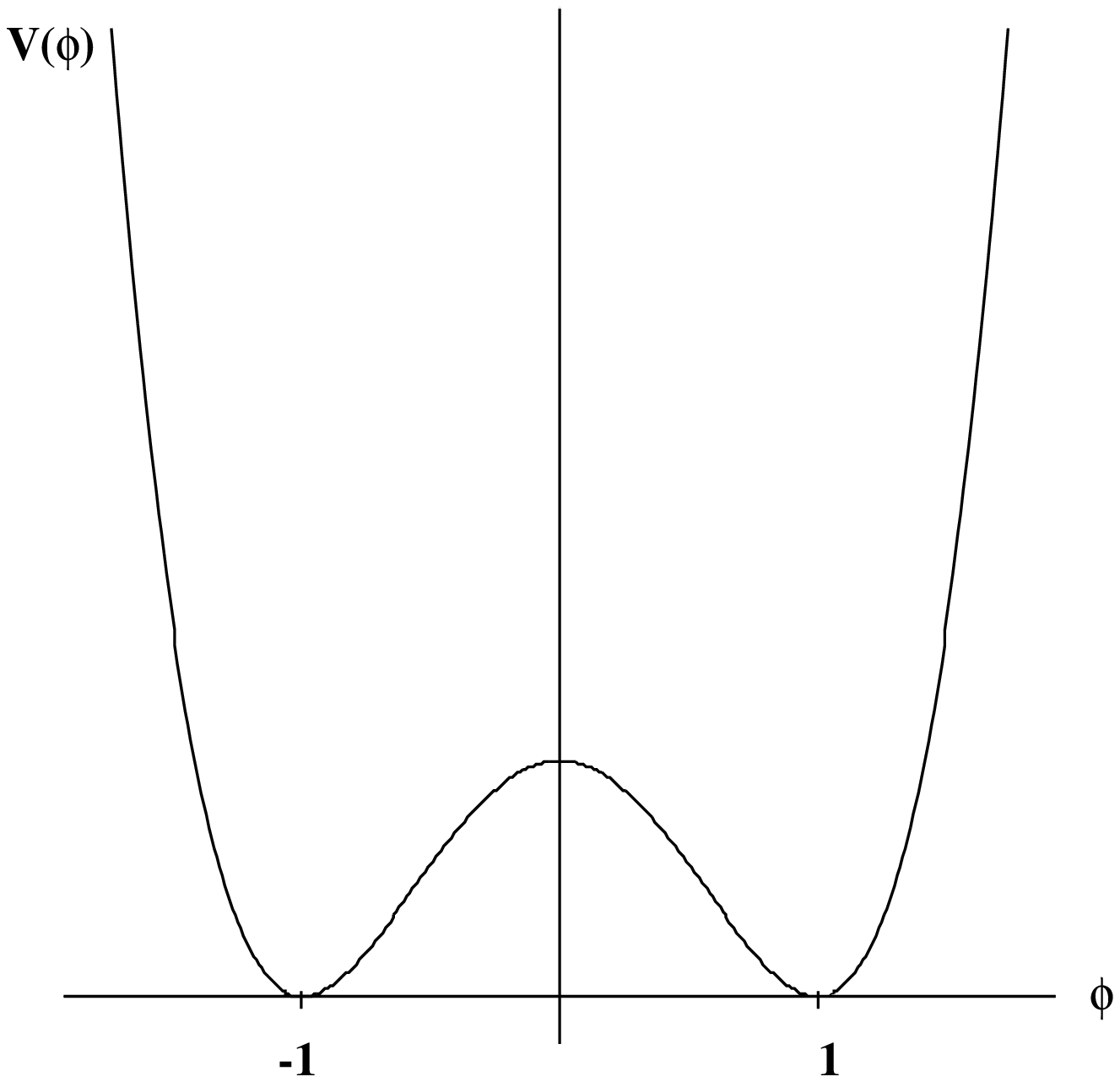,width=7cm}}\\
FIG. 1.	Domain growth potential $V(\phi)$.
\end{center}

Secondly, we shall discuss the spherical version of the 
Sherrington-Kirkpatrick model \cite{Kothjo,Cidi,Cude}
\beqa
 E(s)&=& - \frac{1}{2} \sum_{ij} J_{ij} s_i s_j \nonumber \\
 & & \sum_{i=1}^N s_i^2=N
\label{p2}
\eeqa
where the $J_{ij}$ are quenched random Gaussian variables with zero
mean and variance $1/\sqrt{N}$.  This model shares some, but not all
\cite{Cude} of the properties of `true' mean-field spin-glasses, but
has the advantage that it allows for a complete analytical description.

The third model we shall consider is a `true' spin-glass, in that it
has slow dynamics and aging effects, and its Gibbs measure is given by
a (one step) replica-symmetry breaking Parisi solution.  
It is the $p$-spin version \cite{CrsoI} of the preceding model $p>2$:

\beqa
E(s)&=& - \sum_{i_1<i_2<...<i_p} J_{i_1,...i_p} s_{i_1}... s_{i_p} \nonumber \\
& & \sum_{i=1}^N s_i^2=N
\label{p3}
\eeqa
where the $J_{i_1,...i_p}$ are quenched 
random Gaussian variables with zero mean and variance $p!/2N^{(p-1)}$. 
In the large $N$ limit one can assume that the sum runs over different indices.

In these two last cases we also consider a Langevin dynamics
\beq
\frac{\partial s_i}{\partial t} = - \frac{\delta E}{\delta s_i} - z(t) s_i + 
\eta_i
 \label{p2bis}
\eeq
 where $z(t)$ is a 
Lagrange multiplier enforcing the spherical constraint, and $\eta_i(t)$
are random uncorrelated white noises with variance $2T$. We shall deal
mostly with the zero-temperature case.

Our strategy will be to show that what rather obviously happens in the
first two models also happens in a more hidden way in the third, and
hence argue that such mechanisms are at work also in glassy dynamics.
This will allow us to understand some puzzling aspects of the aging
regime in this kind of system, such as the existence of a
`quasiequilibrium' (FDT) regime of times even in a well out of
equilibrium situation, and the ubiquity of the so-called `marginality
condition' which allows to use pseudo-static methods to obtain certain
dynamical quantities.

The paper is organized as follows.  In Section \ref{sect1} we discuss
some geometric properties of an infinite-dimensional phase space, and
describe how they may lead to a long-time out-of-equilibrium dynamics.
In Section \ref{sect2} we show how these considerations apply to the
ferromagnetic domain-growth case. The Hessian this case
corresponds to a Schr\"{o}dinger problem of 'quantum wires'.
In Section \ref{sect3} we review some results of ref. \cite{Cude} for 
the spherical Sherrington-Kirkpatrick model. 
This model is also very similar to the
domain growth of the $O(N)$ ferromagnet.  A complete study of the
topology of phase-space is extremely simple for it, and in addition we
can get a glimpse at  the effect of non-zero temperature.

Section \ref{sect4} contains the main results of this paper.  We study
there the $p$- spin spherical model ($p>2$), which is `really glassy',
in the sense that its dynamics has an aging regime with long term
memory effects \cite{Cuku} qualitatively close to realistic
spin-glasses.  The equations of motion are in the high temperature
phase  {\em exactly} mode-coupling equations.  The Parisi ansatz for
the replica solution has breaking of the replica symmetry \cite{CrsoI}
and  the  phase-space has exponentially  many valleys
\cite{Kupavi,CrsoII}.  We shall rederive some results of the analytical
solution of \cite{Cuku} on the basis of the present geometrical
scenario, and compare them with the static approach.

\section{Critical points, basins and borders}
\renewcommand{\theequation}{\thesection.\arabic{equation}}
\label{sect1}

{\bf Borders}

Let us start by describing the structure of the  phase-space of a
system with several valleys.  First consider the `critical' or
`stationary' points in which the gradient of the energy vanishes. 
The nature of a critical point is given by the number of negative
eigenvalues of the energy Hessian, which we shall call the `index' $I$
of the point.  The minima (we assume there are at least two) have index
zero, the maxima have index $N$, and the critical points of index one
are the saddle points  connecting two minima.  We shall consider the
rather general situation in which there {\em are} critical points of
every index. We shall denote the `index density' $i \equiv I/N$, 
$0 \leq i \leq 1$.

To each minimum is associated a basin of attraction, defined as the set
of points that will flow through gradient descent to it.  Consider now
the $N-1$ dimensional border of a basin, which we shall  denote it
$\partial_1$. There may be one or several such borders.  Now, a point
that is strictly on a border will never leave the border (by
definition!).  Generically, the trajectory will end in a minimum  over
$\partial_1$ of the energy. Such minima over  $\partial_1$ are
precisely critical points of index one, the saddles separating two true
minima.

Hence, we have that $\partial_1$ is itself divided into basins of
attraction, one for each critical point of index one in it.  Consider
now the $N-2$ dimensional border of one such  basin. We shall label it
$\partial_2$.  Again, a system starting in $\partial_2$ (the border of
the border) will never leave it.  Repeating the argument for
$\partial_2$, we find that it is divided in basins whose minima are the
critical points of index two.  In this way we can iterate the argument
$N$ times, and define $\partial_I$, the border of the border of ...
($I$ times), on which the trajectory generically flows to a saddle
point of index $I$.

All this description may seem rather baroque, given that most points
are not on borders.  However, when we consider an infinite-dimensional
{\em phase} space, the structure of borders becomes relevant for the
following reason:  A random starting point will be contained within a
basin.  Now, since such basin is an $N$-dimensional object, we know
that generically most of its volume is contained within $B \simeq
1/{N}$ of its border $\partial_1$  \cite{Derrida}.  
This in turn means that for $N = \infty$ if the potential is smooth 
enough the system never leaves the vicinity of $\partial_1$ in finite times.

The random point being almost on  $\partial_1$, we can repeat the
argument to find that it will also be very close to a certain
$\partial_2$, ... etc.  We can now iterate this argument a {\em finite}
number of times, to find that the system is near a sequence $
\partial_1,... ,\partial_I$.

We can now understand the origin of the slowing down of the dynamics:  
A system starting strictly on  $\partial_I$ will  end up by being stuck
in a critical point of index $I$. A system starting {\em near}
$\partial_I$ will  be almost, but not completely stuck, and  it slows
down. 
For long times we have that the trajectory manages to distance itself from
$\partial_I$ corresponding to degrees of `bordism' $I$ that are smaller
and smaller but still $i=O(1)$ never distancing itself from
$\partial_I$ corresponding to finite $I$.
In other words, the neighborhoods of the critical points of $i
\simeq 0$ are for long but finite times efficient in trapping the system.
Figure 2 shows how this would come about in a two dimensional phase-space: 
points starting near the borders have trajectories that take long to reach 
the minimum (of course, the condition of starting near the border is imposed
in two dimensions, while it arises naturally in many).

\begin{center}
   \parbox{8.5cm}{\epsfig{file=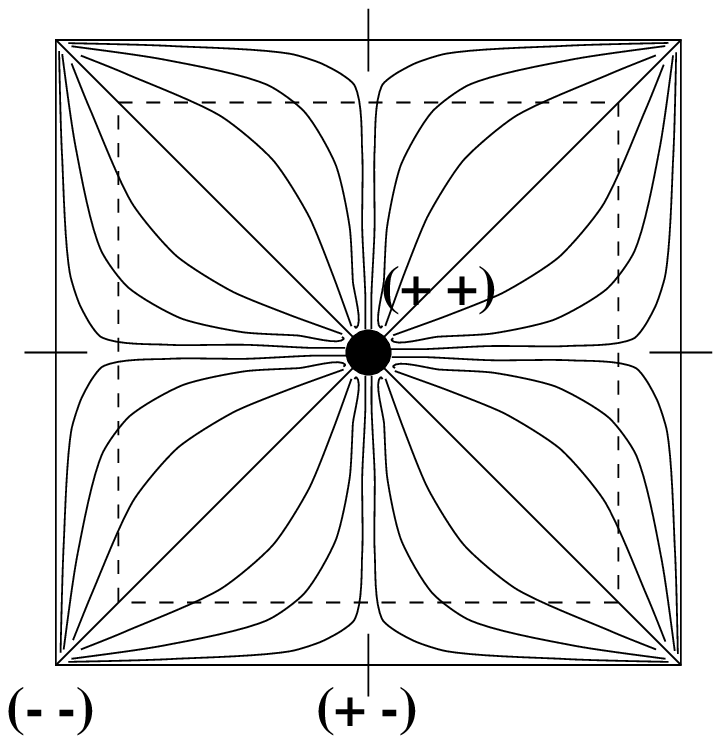,width=8cm}}\\
FIG. 2.	A schematic representation of a basin in $2D$ phase-space.
	The signs indicate the indices `$I$' of the critical points 
	(maxima are vertices and the minimum is at the center of the square).
 	The trajectories starting near an edge take longer to fall. The
	dotted line represents schematically the 'border'.
\end{center}

What we have described is a non-Arrhenius mechanism for aging which  works
even at zero temperature, and which does not involve any sudden
processes of barrier jumping.  This mechanism, as we shall see below, is
at work in the case of domain growth; the important question here is
that it seems rather generic for systems with well separated minima,
whether we are able to identify a spatial structure  for them or not.

{\bf Hessian}

At long times one knows
that the gradient must be small (because the system has slown down),
but is still non-zero so that we are not precisely in a critical point.
Indeed, at zero temperature:
\beq
 |\nabla E|^2= -\frac{dE(t)}{dt} \;\;\; \rightarrow \; 0
\eeq

If we now consider the matrix of second derivatives of the energy,
\beq
{\bf H}_{ij} =  \frac{\partial^2 E}{\partial s_i \partial s_j}
\label{hessian}
\eeq
we know that in a neighborhood of a critical point of index $I$ the
Hessian has  $I$ negative eigenvalues.  A natural  assumption
suggested by the scenario described above is that 
if we follow a trajectory that starts near $\partial_I$ the spectrum of the 
Hessian will  be similar at every time to that of a nearby critical point,
reflecting the degree of `near bordism' at that time.
This is easy to see in Fig.2, a trajectory starting near the border will
have typically one positive and one negative eigenvalue, until it
`unsticks' from the border and it ends by having two positive eigenvalues.

The dynamics at long times will be such that the $H$ will have a
distribution of eigenvalues $\lambda_\mu$ containing $R(t)$ negative
eigenvalues, with $R(t)$ decreasing with time, corresponding
to a situation in which the landscape at time $t$ is similar to the
landscape at a nearby critical point of index $I=R(t)$.
The density of the eigenvalues of the $H$ at time $t$,
$\rho_t(\lambda)$ will then contain a bulk of positive eigenvalues,
plus a tail extending down to some small negative minimal eigenvalue.
The integral over the tail of negative eigenvalues is  $R(t)$.
The precise manner in which $\rho_t(\lambda)$ tends to its limit
$\rho_\infty(\lambda)$ is model-dependent, we shall describe them in
detail below for the three models discussed in this paper. The main
features are, however, the same: a distribution over positive $\lambda$
that stabilizes quickly, plus a tail that extends up to negative
eigenvalues which tends to disappear slowly with time.

Let us see that  the velocity vector points, for long times,  in the
directions of low (positive and negative) eigenvalues of the Hessian.
If there is slower than exponential decay of the energy, we have that
\beq
 \frac{ (\nabla E)^+ {\bf H} (\nabla E)}{|\nabla E|^2} = 
\frac{d^2 E}{dt^2}/\frac{d E}{dt} \;\; \rightarrow 0
\eeq
Denoting $v_\mu$
the component of the velocity in the direction of the eigenvalue 
$\lambda_\mu$ of ${\bf H}$, this means that:
\beq
\frac{\sum_\mu v^2_\mu \lambda_\mu}{\sum_\mu v_\mu^2} \;\; \; \rightarrow 0
\eeq

Hence, we have that at long times, the particle moves in a gorge 
with locally many directions in which it is a minimum,  plus a few
almost flat  directions whit positive and negative curvatures: the
system is `critical' or `marginal' at all finite times. The gradient
is small, and is pointing along the almost-flat subspace. We have
arrived at this picture by arguing that the dynamics takes place along
ridges, and we now find that remarkably, in high dimensions, a ridge
can behave also as a channel.
It is important to remark that the claim here is not that at long times 
there should be slow degrees of freedom (this is obvious), but that
the existence of such slow directions is a natural consequence of the
dominance of borders in high dimensionalities.

{\bf Energy differences between critical points, speed of descent.}

Let us now review a few results that will be useful in the discussion that 
will follow.

Given two critical points of index $I$ and $I+1$, respectively, which
are joined by a gradient line (e.g. a minimum and a saddle), we want to
estimate what the  energy difference of such a `step' may be. Consider
a `ladder' of such steps, taking us from a minimum to a saddle, from a
saddle to a critical point of index $I=2$, and so on up to a maximum.
If the system does not have infinite energy density configurations, the
total energy climbed is $O(N)$, in $N$ steps.  Hence, at most a finite
number of such steps can be of $O(N)$, and there must be steps of $O(1)$. 
This means that most of these steps are almost flat.

For $D$ dimensional systems with short range interactions we can say
more. By considering a domain of a phase A growing against a domain of
another phase B, one finds that:

i) If there is a minimum whose energy is $O(N)$ above another
minimum, the energy of the barrier separating them is $O(1)$
\cite{Pa}.

ii) Between two minima there is at least one saddle (index $I=1$)
that is at most $O(N^{1-1/D})$ in energy above them.

Let us now give an upper bound for the time of descent between two
points.  Consider a point in phase-space $\vec s^a$, and another point
$\vec s^b$ which is downhill from  $\vec s^a$ along a gradient line.
Let their energy difference (energy density difference) be $\Delta
E_{ab}$ ($\Delta e_{ab} \equiv \Delta E_{ab} / N$), and  distance
$d_{ab}= |\vec s^a -\vec s^b|$.

We now ask ourselves what is the minimal possible time for descent from
$a$ to $b$.  It is easy to show that the time is minimal if the path
joining $a$ and $b$ has constant gradient $= \Delta E_{ab}/d_{ab}$.
Hence we have:  \beq t_{a \rightarrow b} \geq \frac{d_{ab}^2}{\Delta
E_{ab}} = \frac{B(a,b)}{\Delta e_{ab}} \label{inua} \eeq Consider now
two minima and their associated saddle point ($I=1$), and let each
minimum be well separated from their common saddle $B(saddle,minimum)>0$.
An immediate consequence of (\ref{inua}) is that if the energies 
differences between barriers and minima scale with
the system size slower than $N$, the time for descent from the
neighborhood of the saddle to the neighborhood of either minimum is
infinite.  This argument certainly holds for finite dimensional systems
with short range interactions, since for them $\Delta e (barrier,
minimum) \leq O(N^{-1/D})$.

Furthermore, since as we have seen before the gradient lines joining 
most critical points have energy differences of order smaller than $N$, 
we find that if two such critical points are well-separated 
the time of descent from neighborhoods of each is again infinite.

{\bf Ergodicity}

Having claimed that the motion takes place near borders, 
we must reconsider what is it that we shall understand by
`ergodic component' in an out-of equilibrium situation.
Suppose we call `ergodic component at time $t$' the connected
set of points that includes the configuration $\vec s(t)$
and  have an energy lower or equal than $E(t)$ \cite{Siho},
i.e.  the set of points to which the system can be driven without work.

Let us now argue that an ergodic component so defined includes at any
finite time  many, and in systems with finite spatial dimensions,  
{\em all} minima:  At time $t$, there are within the ergodic component
several points of index $I=1$, having various energies. Each time
$E(t)$ reaches the energy of one of these points, the constant-energy
surface develops a separatrix and there is a disconnection of a subset
of the ergodic component.

It may happen that many (or even all) the critical points of index one
are at an  energy of order smaller than $N$ above the minima. Indeed,
this will be always the case with  finite spatial dimensions and short
range interactions. Now,  the excess energy $E(t)-E(t=\infty)$ is of
$O(N)$ at any finite time. In that case the ergodic component never
disconnects and it includes {\em all} the minima at any finite time
$t$:  the dynamics is such that the system refuses to break its
ergodicity at finite times.

At non-zero temperature we can discuss ergodicity at a given time from
a related but different point of view by asking ourselves whether a
configuration at a given time $t$ is doomed to fall in an assigned
state, or it may change basin due to  thermal fluctuations at times
$>t$. That is, we are asking if the `target' state is fully determined
by the configuration at time $t$.

We cannot answer this question in general, but in section \ref{sect3}
we will show in a particular model that there is at any given finite
time a non-zero probability of changing basin - and this long before
the system has had time to cross barriers between minima.  One can
suspect that this is quite general, given that at any finite time the
system has descended very little from the ridge separating basins (it
is close to $\partial_1$) so the thermal fluctuations may well make it
jump across the ridge and head for a different state.

{\bf `Quasi-equilibrium' regime and marginality}

A rather surprising feature that appears in spin glasses, is that if
one observes  the correlation and response functions at two long but
not very separated times, they depend on time-differences and obey the
fluctuation-dissipation theorem (FDT), just as in a system in
equilibrium - even if the system is visiting a region of phase-space to
which it will never return.

This can be understood within the scenario described above:  the fast
relaxations  are dominated by the local directions with large second
derivatives. The  form of $\rho_t (\lambda)$ for large times determines
the precise time-dependence of these relaxations.  The slow drift
phenomena are related to the motion along the almost flat subspace, i.e.
the tail of $\rho(\lambda)$ for $\lambda$ around zero.  The fact that
the `quasi-equilibrium' correlations and response functions depend on
time-differences reflects the fact that the form of  $\rho_t (\lambda)$
for $\lambda$ well above zero stabilizes  quickly, the `channel walls'
in most directions preserve their form.

Another surprising  question in mean-field spin-glass dynamics is the
so-called `marginality condition'.  In its original form \cite{Hofr},
the marginality `principle' stated that the dynamical values of
energy, susceptibility, and the so-called `anomaly', are determined by
the requirement that the fast relaxations (in the FDT regime) be
`critical' or `marginal', in the sense that they follow power laws
instead of exponentials.  The dynamics considered there was made
manifestly out of equilibrium (though not aging) by making the
Hamiltonian itself (slowly) time-dependent.  Because in many models the
dynamics in a true equilibrium state is non-critical, the results so
obtained  differ from those at equilibrium for them.

It also turned out (although no general proof exists at the moment),
that one can obtain the large-time limit of  some one-time quantities
by solving a static problem and imposing the solution to have marginal
stability \cite{Hofr,Cuku}.

The question seems very puzzling: why should the system always choose
to fall in a state that is marginal, refusing to see those that are
not?  Within  the present geometrical scenario, the question is quite
clear:  the dynamics is by construction non-equilibrium, at least at
the beginning, even if we always consider a time-independent
Hamiltonian.  Then we argue that the system never achieves (even local)
equilibrium, it doesn't fall {\em anywhere}, but  is confined near
borders of the basins and the Hessian at long times contains a
(decreasing) number of negative eigenvalues.  In this sense the
dynamics is automatically at all times marginal, whatever the stability
of the true minima.

If the minima are $O(N)$ below the borders, we will then observe a finite 
energy-density difference with respect to them at any finite time. 
This last thing cannot happen in finite-dimensional
systems, but it does happen in the mean-field model we shall discuss 
in section \ref{sect4}.  
In that section we shall discuss this question in more detail.

The origin  of a `quasi-equilibrium' regime and the marginality of the
long time dynamics are  easy to understand in the case of ordinary
domain growth:  The  response and the correlation function at small
time-differences are dominated by the bulk of the domains, which are
locally (in real space) in equilibrium.  The marginality of dynamics is
given by the zero modes associated with moving a domain wall.  Again,
the main point here is that by considering the phase space geometry we
can understand why these things happen in systems which either do not
have a real space domain structure, or of whose real space structure we
do not know.

\section{Domain Growth}
\renewcommand{\theequation}{\thesection.\arabic{equation}}
\label{sect2}

Let us see how the description in the preceding section applies to
the case of ferromagnetic domain-growth \cite{Bray}, 
eq. (\ref{??}). For definiteness we restrict ourselves
to two dimensions. 

We denote the size of the system $V=L^2$, and  $L \rightarrow \infty$.
This case is somewhat complicated by the fact that there is
translational invariance, and hence the discussion has to be done
modulo translations.

The model  has  two zero-energy ground states $\phi(x)=\pm 1$, which we
depict in Fig3.b  in black and white, respectively.  For long times,
the system consists of domains of the two types separated by sharp
domain walls (Fig3.a).  The energy over the minima is at long times
proportional to the total length of all domain walls, which is at
finite times $O(L^2)$.

The phase-space square distance to the $\pm$ minima is given by:
\beq
B(\phi,\pm)=\frac{1}{V}\left[\int d^D \phi^2(x) +
1 \pm 2 \int d^D x \phi(x)\right] 
\eeq
At time $t$ the typical domain size is $r(t)$, and by hypothesis we are
in the regime $r(t)<< L$ (Fig3.a), so that $B(\phi,\pm)$ is $\simeq 2$
at all finite times. The system clearly remains far from either minimum.

The saddles separating  minima are easily constructed:  they correspond
(Fig3.c) to dividing the volume in two equal pieces of opposite phases,
with  straight interfaces.  There are two  continua of such saddles,
obtained by translation and $90^o$ rotation.  Their energy is $O(L)$,
so that at any finite time the energy of the system is way above the
energy required to go from one basin to the other.

\begin{center}
   \parbox{8.5cm}{\epsfig{file=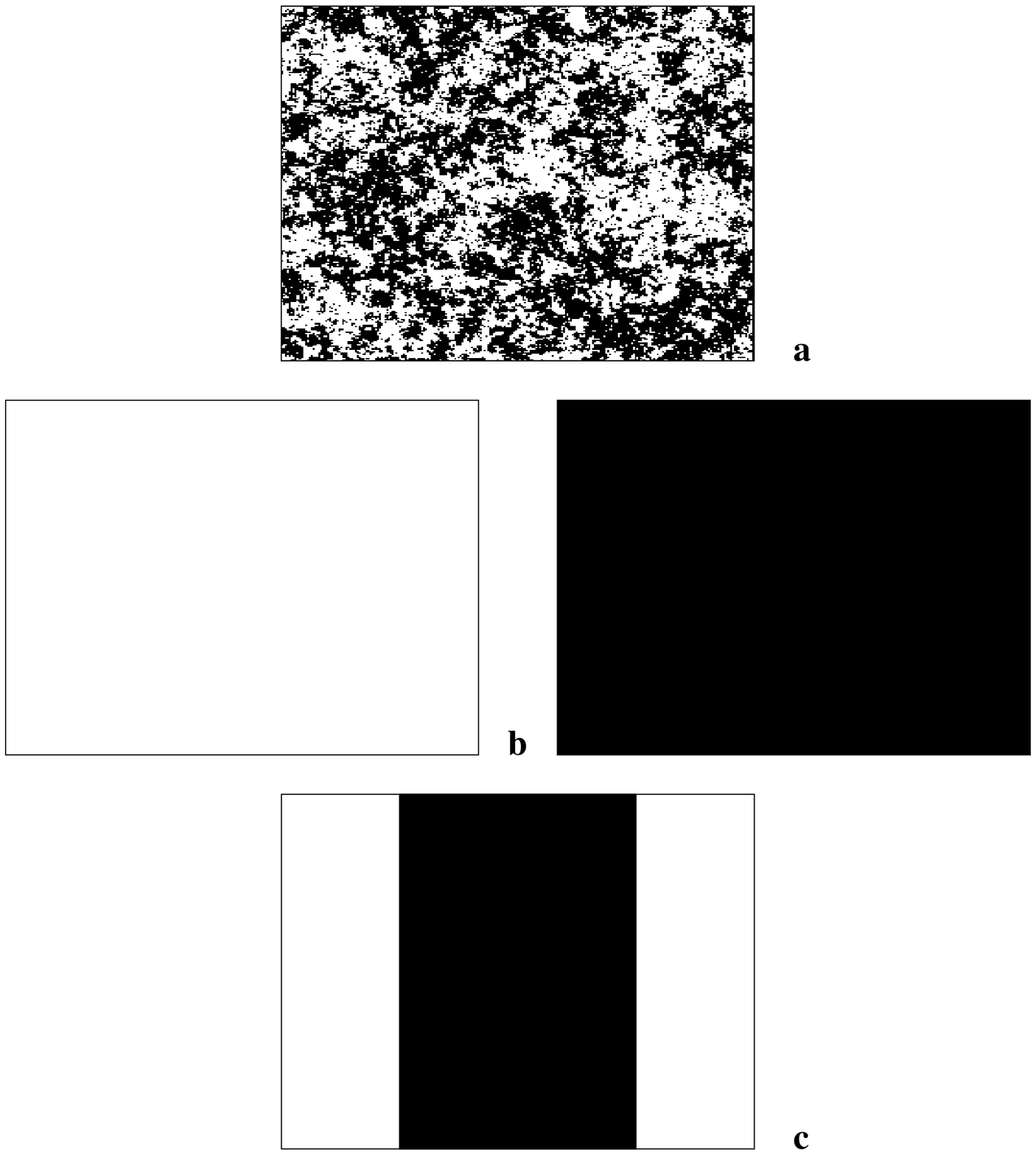,width=8cm}}\\
   	FIG. 3.
     a) a domain configuration\\
     b) the two states $\phi=-1$ and $\phi=+1$\\ 
     c) a saddle configuration.
\end{center}

The Hessian matrix in the  phase-space point $\phi(x)$ is the operator:
\beq
H=\frac{\delta^2 E}{\delta \phi(x) \delta \phi(y)} = 
[-\nabla_x^2 + V''(\phi(x))]\delta(x-y)
\label{h1} 
\eeq 
The eigenvalues $\lambda_\mu$ and eigenvectors $\psi_\mu$ of the
Hessian evaluated in $\phi$ are then obtained from the Schr\"{o}dinger
problem:
\beq
 [-\nabla_x^2 + V''(\phi(x))] \psi_\mu(x) = \lambda_\mu \psi_\mu(x)
\label{shroe}
\eeq
with `potential energy' given by $V''(\phi(x))$, and periodic boundary
conditions for the wave functions.  The Schr\"{o}dinger potential  is a
well that follows the domain walls and  rapidly tends to  $\sim V''(\pm 1)$
away from them.  Figure 4.b shows the Schr\"{o}dinger potential
across a domain wall.

In order to obtain the complete spectrum, let us first consider a
single, straight wall.  Translational invariance tells us that
\beq
\psi_0(x)= \phi'(x)
\label{bound}
\eeq
is a bound eigenvector  of the Schr\"{o}dinger potential with 
$\lambda \sim 0$, which corresponds to shifting the wall. 
In the saddle point configuration of Fig3.c (where two domain walls are 
present), we have that the $\lambda = 0$ eigenvalue is precisely $\phi'(x)$,
which is an {\em odd} function localized near the two domain walls.
Since this function has a node, there must be a lower eigenvector which is
similarly localized but {\em even}:  its eigenvalue is then negative,
and it corresponds to moving the two domain walls in opposite directions.

We can now discuss the structure of the Hessian at  large times.
The Schr\"{o}dinger potential consists then of thin wells that follow 
the domain walls.
The  structure of bound eigenvalues of such a problem can be
appreciated easily by noting that it corresponds to a problem of
`quantum wires' (a `wire' being the region of each domain wall), a
problem of localization that has been extensively studied in the
literature \cite{martorell}.

\begin{center}
  \parbox{8cm}{\epsfig{file=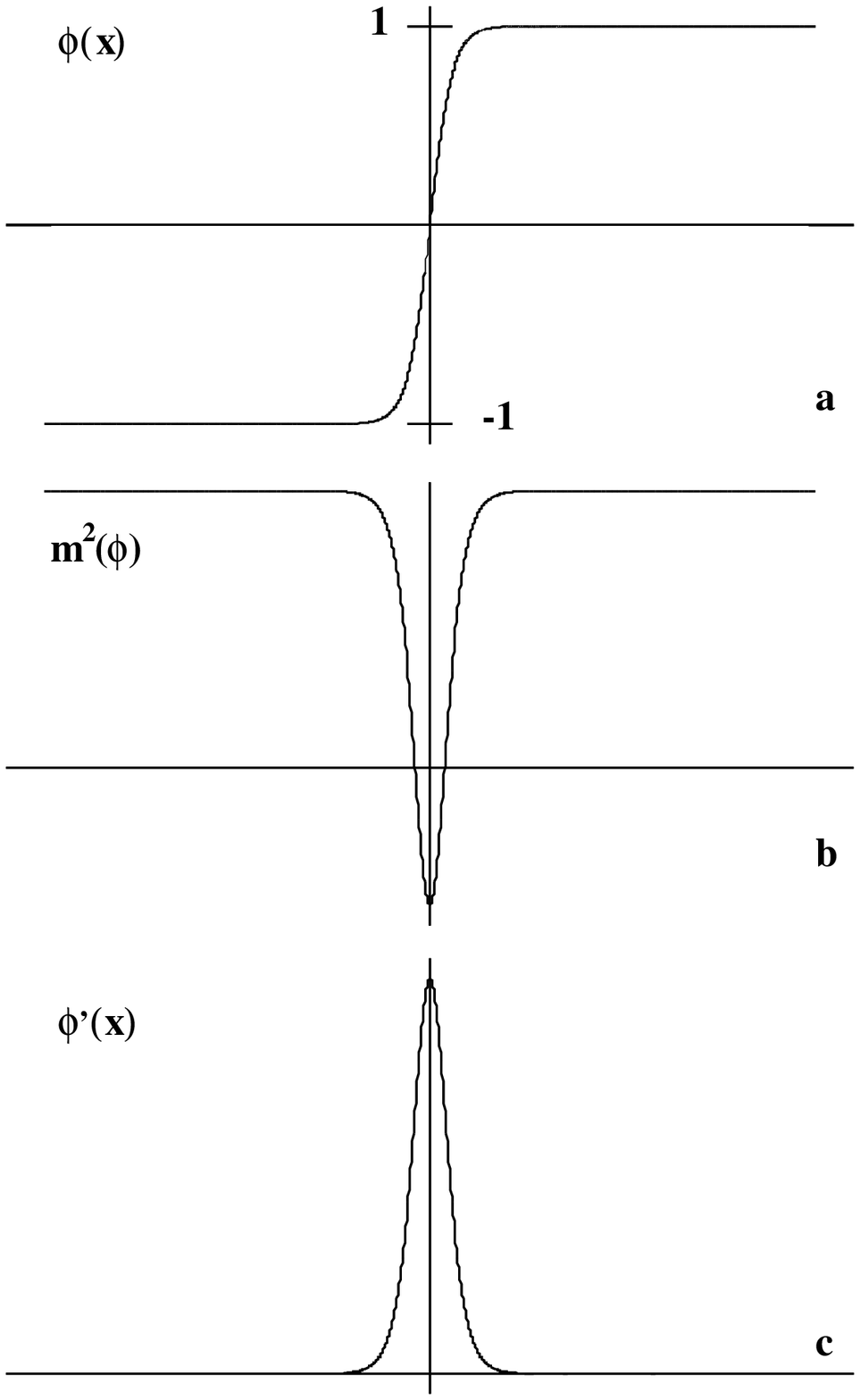,width=7cm}}\\
FIG. 4. 
	a) $\phi(x)$ across a domain wall\\
	b) The Schr\"{o}dinger potential $V"(\phi)=m^2(\phi)$  
	 of the corresponding domain wall\\
	c) Schr\"{o}dinger wave function $\psi_{\lambda}=\phi'(x)$ with
        $\lambda \sim 0$ across the domain wall.
\end{center}

The eigenvectors of $H$  fall into three classes:

i) all eigenvectors with $\lambda_\mu > V''(\pm 1)$ are unbound. They
are simply the bulk oscillations of the magnetizations, and are little
affected by the domain structure.

ii) there are the bound eigenvectors which in the direction
perpendicular to the walls of the domains are  essentially like
(\ref{bound})  (Fig4.c), and oscillate like $e^{ikw}$ in the direction
$w$ along  the walls.  Their eigenvalues are proportional to
$k^2$, and they correspond to the massless spectrum of fluctuations (of
length $1/k$) of the domain walls.

iii) finally there are negative eigenvalues localized \cite{martorell}
in the more curved regions of the domain wall, with eigenvalues
$\lambda \sim -1/r^2$, where $r$ is the local curvature of the domain
wall in the region of localization (these are the localized states of
the quantum wire problem).

At any finite time there is then in addition to the  `bulk' of large
eigenvalues (i), a tail of small, positive (ii) and negative (iii)
eigenvalues.  Clearly, as time passes all local curvatures become
smaller and the domain walls become more and more sparse, so the
negative eigenvalues tend to approach zero and simultaneously the
distribution of eigenvalues contains less and less eigenvalues smaller
than $\lambda_\mu <  V''(\pm 1)$.

The `velocity vector'  $\frac{\partial \phi}{\partial t}$ is in real
space concentrated along the regions of fastest variations of $\phi$,
i.e. in functional space it mainly points in the direction of (ii) and
especially (iii).

The fast response  is dominated, at large times, by the eigenvalues
 $\lambda_\mu > V''(\pm 1)$, and coincides with the one obtained at
 equilibrium.

\section{Spherical SK model}
\renewcommand{\theequation}{\thesection.\arabic{equation}}
\label{sect3}

The spherical $p=2$ model has been introduced in \cite{Kothjo}, where
it was shown that its statics has a low-temperature phase with two
states. The replica solution does not have replica symmetry breaking.
The long time dynamics \cite{Cidi,Cude} has two regimes of times:

i) for $t,t' \; \rightarrow \infty$ and $t-t'$ finite the correlation
and response functions are time-translational invariant, and satisfy
FDT:
\beqa
C(t,t')&=& C_{FDT}(t-t') \nonumber \\
R(t,t') &=& R_{FDT}(t-t') = 
       \frac{1}{T}\frac{\partial  C_{FDT}(t-t') }{\partial t'}
\label{fdt1}
\eeqa
with $C(t,t)=1$

For large $t-t'$, but still $<<t$, 
\beq
C_{FDT} \; \simeq \;  q_{EA}+ A (t-t')^{-1/2} \;\;\; ; \;\;\; R_{FDT} 
\propto (t-t')^{-3/2}
\eeq
where $q_{EA}= 1-T$ is the Edwards-Anderson parameter.  At strictly
zero temperature $q_{EA}=1$, so that in order to see something of the
decay of the correlation, and to check FDT, in this regime we will have
to go to small but non-zero temperature.

ii) The system is not, however in equilibrium at any finite time:  In
the regime of large, comparable, times in which $t,t' \; \rightarrow \;
\infty$ and  $0<t'/t < 1$ the correlation function is a non-homogeneous
function of $t'/t$, while FDT is violated (see ref. \cite{Cude}).

Let us see how the considerations of section \ref{sect2} apply to this
case.  The energy of the  model, defined by the Hamiltonian (\ref{p2})
reads, in the basis $s_\mu$ in which the matrix $J_{ij}$ is diagonal:
\beq
E= - \frac{1}{2} \sum_\mu J_\mu s_\mu^2
\eeq
with the spherical constraint 
\beq
\sum_\mu  s_\mu^2 = N
\eeq
where $J_\mu$ are the eigenvalues of $J_{ij}$, which for large $N$ are
distributed with a semicircle law with support $(-2,2)$ \cite{Me}.  
Let us denote $s_1,s_2,...,s_N$ the directions associated with the 
eigenvalues in decreasing order ($J_1 \simeq 2, ..., J_N \simeq -2$).

The stationary points of $E$, when restricted to the sphere, are the
directions of the eigenvectors. There are two minima $s_\mu=\pm {\sqrt
N}\delta_{\mu,1}$, two saddle points separating them $
s_\mu=\pm {\sqrt N} \delta_{\mu,2}$,  and in general two critical
points of index $I$: $s_\mu=\pm \sqrt{N} \delta_{\mu,I}$. 
The energy difference between the minima and the saddles is easily
shown using the semicircle law to be of $O(N^{1/3})$.

The equation of motion in terms of these variables is, at zero temperature:
\beq
\frac{\partial s_\mu}{\partial t} = (J_\mu - z(t)) \; s_\mu(t) 
\eeq
We assume that the initial configuration  ${\vec s}^{\ 0}$ is uncorrelated
with the potential, and  hence  in the eigenbasis of $J_{ij}$ each
$s_\mu^0$ is a random number of $O(1)$.  The solution to the equations
of motion then is:
\beq
s_\mu(t)=s_\mu^0 \; \exp\left [ \int_0^t  (J_\mu-z(\tau)) d \tau  \right]
\label{sol}
\eeq
where 
\beqa
z(t)&=& \sum_\mu J_\mu s_\mu^2 = -2 e(t) \;\; < 2 \;\; 
\forall \;\; t \nonumber\\
z(t) &\simeq& 2- \frac{3}{4t} \;\; as \;\; t \rightarrow \infty
\label{zeq}
\eeqa
From (\ref{sol},\ref{zeq}) one sees that $s_1(t)$ does not change sign
and its absolute
value grows steadily. Hence, the two basins of attraction are the set
of points:
\beqa
\vec s \;\; &/& \;\; s_1>0 \nonumber \\
\vec s \;\; &/& \;\; s_1<0
\eeqa

The border $\partial_1$ is then the   set $\partial_1=\{ \vec s \;/\;
s_1=0\}$.  Repeating again the argument, we conclude that $\partial_1$
is itself divided into to basins leading to the two saddles. The border
between these is $\partial_2=\{\vec s \; / \; s_1=0  \; , \;s_2=0 \}$.
In general $\partial_k = \{\vec s \; / \; s_1=0,...,s_k=0 \}$.

The normalized squared distance to $\partial_1$ is $B(\partial_1,\vec
s(t))=s_1(t)^2/N$, and remains of $O(1/N)$ at all finite times. In
general, this is true for the distance to $\partial_I$ ($I$ finite),
given by:
\beq
 B(\partial_I,\vec s(t))= \frac{1}{N}\sum_{k=1}^{I} s_k(t)^2
\eeq

We now turn to the study of the Hessian. Because the system is
spherically constrained, we have to restrict it to the directions along
the sphere.  A direct way to see which is the relevant operator is to
consider the dynamics of two trajectories that are close to one
another: $\vec s (t)$ and $\vec s (t)+\vec \sigma(t)$, with $|\vec
\sigma|$ small. Using the equations of motion and the constraint, plus
the fact that
\beq 
\vec s (t) . \vec \sigma(t) =0
\label{constr}
\eeq
 We have, in the original basis:
\beq
\frac{\partial \sigma_i}{\partial t} = -H_{ij} \sigma_j - \frac{1}{N} \sum_j 
\frac{ds_j}{dt} s_i \sigma_j
\label{ccc}
\eeq
where
\beq
H_{ij}= 
\frac{\partial^2 E}{\partial s_i \partial s_j} + z(t) \delta_{ij}
\label{mmf}
\eeq
is the effective Hessian for a spherically constrained system. 
The last term in (\ref{ccc}) serves to impose
the preservation of the constraint, and tends to zero with time.

For this model the Hessian reads:
\beq
{\bf H}_{ij}=-J_{ij}+z(t) \delta_{ij}
\label{hes}
\eeq
 We
note that $H$ is time-dependent (through $z$), but its eigenbasis does
not depend on time.

We now know the structure of eigenvalues of the Hessian for all times:
$\rho_t(\lambda)$ is a shifted semicircle law with support in the
interval $[-2+z(t),2+z(t)]$.  The  distribution is, for large times, 
a semicircle starting in
\beq
\lambda_{min}(t)= -\frac{3}{4t}
\eeq
and extending up to $\lambda \simeq 4$.
The number of negative eigenvalues goes as $ \sim N t^{-3/2}$.
The limiting form of the distribution, $\rho_{\infty}(\lambda)$
is a semicircle with support in $[0,4]$.

Let us now discuss how the long-time structure of the Hessian is
reflected in the response and correlation functions.  Consider the
effect of a small kick on the system at time  $t_w$, in the direction
of the magnetic field of intensity $\Delta h$ and duration $\Delta t$.
It will shift the configuration $s_i(t_w)$ to $s_i(t_w)+ \sigma_i(t_w)$,
where $\sigma_i(t_w)=\Delta h \; \Delta t $.  
The response function at subsequent times is  the increase of magnetization
due to the field, per unit of  $\Delta h \; \Delta t$, i.e.:
\beq
R(t,t_w)= \frac{1}{\Delta h \; \Delta t \; N} \sum_i \sigma_i(t)=
 \frac{\sum_i \sigma_i(t) \sigma_i(t_w)}{\sum_j \sigma_j(t_w)^2}
\eeq
We can now solve eq. (\ref{ccc}), neglecting its last term, to get:
\beq
\sigma_i(t)=\sum_j \left[ e^{-\int_{t_w}^t {\bf H} (\tau)d \tau} \right]_{ij} 
\sigma_j(t_w) 
\eeq
Leading to:
\beq
R(t,t_w)=\frac{1}{\sum_j \sigma_j(t_w)^2} \sum_{ij} \sigma_i(t_w) 
\left[ e^{-\int_{t_w}^t {\bf H} (\tau)d \tau} \right]_{ij} \sigma_j(t_w) 
\label{3.17}
\eeq 
Because the eigenbasis of the Hessian is uncorrelated with the direction
of the magnetic field, this becomes, in the large $N$ limit:
\beqa
R(t,t_{w})&=& \frac{1}{N} \; tr_\perp \; \left[ 
e^{-\int_{t_w}^t {\bf H} (\tau)d \tau} \right] \nonumber\\
& & =  tr_\perp \; \left[ e^{- \; <{\bf H}>_{t,t_w} \; (t-t_w)} \right]
\label{3.18}
\eeqa
where the $tr_\perp$ denotes the trace restricted to the directions 
tangential to the sphere, and we have defined  the  time-averaged 
Hessian as
\beq
 < H_{i,j}>_{t,t_w} \equiv \frac{1}{t-t_w} 
\int_{t_w}^t H_{ij}(\tau) d \tau 
\label{3.19}
\eeq
It is clear that in (\ref{3.18})  the contribution of  the tail of low
eigenvalues of the averaged Hessian is negligible for finite
time-separations $t-t_w <<t$, and that we can for these time
separations substitute the averaged Hessian by the asymptotic
(semicircle) distribution $\rho_\infty(\lambda)$.  Hence, we find that
the fact that $\rho_t(\lambda)$ has a limit implies time-homogeneity in
this regime of times.  Furthermore, the fact that
$\rho_{\infty}(\lambda) \propto \lambda^{1/2}$ for small $\lambda$ (but
large compared with $1/t$) implies that $R(t-t_w) \propto
(t-t_w)^{-3/2}$.

For time-separations of $t-t_w$ of the order of $t$, the exponential in
(\ref{3.18}) selects the tail of lowest eigenvalues, and the FDT
regime breaks down: the tail of almost flat directions of ${\bf H}$ is
responsible for the `aging regime' $t-t_w = O(t)$.

For this simple model, the calculation can be carried out explicitly,
using the asymptotic form for $H$ at long times:  
\beqa 
<H_{ij}>_{t,t_w} (t-t_w)&=&  -J_{ij} (t-t_w)+ 
\left[ \int_{t_w}^{t} z(\tau) d \tau \right] \; \delta_{ij}  \nonumber\\ 
&=&[-J_{ij} + 2 \delta_{ij}](t-t_w) - \frac{3}{4} \  \delta_{ij} \ln 
\frac{t}{t_w} 
\label{3.20}
\eeqa 
from which an expression for the aging regime can be
readily found.

Let us now turn to the `fast' correlation function  at small but non-zero 
temperature and at two large but not very separated times.
Because the motion along the flat directions is slow,  it
can be neglected for short time differences. 
On the other extreme, one can assume that the system is equilibrated in the 
`fast' degrees of freedom corresponding to large eigenvalues of ${\bf H}$.
Since, as we have seen, in this regime of times it is  only such degrees of 
freedom that also contribute to the response function, we conclude that FDT 
must hold. The correlation function then reads:
\beqa
C(t,t')&=& 1- T \int_{t'}^{t} R(t,t'') d \tau \nonumber \\
&=& 1- T \;  tr_{\perp} \int_{t'}^{t} \; dt'' 
 \left[
e^{-\int_{t''}^t {\bf H} (\tau)d \tau} \right]
\label{corr1}
\eeqa
Since $t-t'$ is by assumption $<<t$, we can neglect the variation of 
$\bf H$ to get:
\beqa
C(t,t')&=& 1- T \; tr_\perp \left\{ {\bf H}^{-1}(t) 
\left[ 1- e^{-{\bf H}(t) \; (t-t')} \right] \right\} \nonumber \\
&\simeq& 1-  T \sum_{\mu} \frac{1- e^{-\lambda_\mu (t-t')}}{\lambda_\mu}
\label{corr2}
\eeqa

We can now check that the assumptions we made above are consistent: 
for $t-t'$ finite and large $t$, the numerator in (\ref{corr1}) acts 
as a low $\lambda_\mu$ cutoff: the (few) positive and negative  $\lambda_\mu$ 
that are close to zero $O(1/t)$ do not contribute. 
As $t-t'$ becomes comparable with $t$ the approximation (and hence the validity 
of FDT) breaks down because on the one hand we can no longer suppose the 
constancy of ${\bf H}$, and on the other hand the slow degrees of freedom 
($\lambda_\mu \sim 0$) start contributing and we cannot assume that they are
equilibrated.

The inverse eigenvalue $1/\lambda_\mu$ is the typical length of the 
fluctuations in the direction $\mu$, it is infinite if $\lambda \leq 0$. 
The Edwards-Anderson parameter is:
\beqa
q_{EA}= \lim_{t-t' \rightarrow \infty} \lim_{t \rightarrow \infty} C(t,t')
\eeqa
The quantity
\beq
a_q \equiv \frac{1-q_{EA}}{T}
\label{aq}
\eeq
precisely measures the average width of the channel, the order of the limit 
ensures (via the numerator in (\ref{corr2})) a cutoff in the
directions in which the system is unbound, i.e.  it selects the `walls' 
of the channel against the longitudinal direction.
Note that in this model there is {\em no} discontinuous process of escape 
from a trap.

Finally, let us discuss the question of ergodicity breaking.  
As we have noted already, the separation between minima is of order
$O(N^{1/3})$, while the energy above the minima is, for large times
$\Delta E (t) \simeq \frac{3}{8t}N$, well above the barrier.  In this
model we can also ask ourselves about ergodicity in the other sense of
section \ref{sect2}: whether there is the possibility at non-zero
temperature, of changing basin spontaneously.

We have seen that it is the sign of $s_1$ (in the eigenbasis of
$J_{ij}$) which defines the basin.  The evolution of $s_1$ is, for
finite temperature $T$ given by (see II.4 of ref. \cite{Cude}):
\beqa
s_1(t)&=& s_1 (t=0) e^{-\int_0^t d\tau (z(\tau)-2) d \tau} \nonumber\\
& & +\int_0^t dt'' e^{-\int_{t''}^t d \tau (z(\tau)-2)} \eta(t'')
\label{3.21}
\eeqa
The first term is deterministic, while the second is a Gaussian random
variable with variance:
\beq
2T \int_0^t dt'' 
  e^{-\int_{t''}^t  2(z(\tau)-2) d \tau} 
\label{3.22}
\eeq
a quantity of order one. Hence, there is at any finite time the
probability that $s_1(t)$ will change sign (unless, of course, the
system started well-within a basin: $s_1(t=0)=O({\sqrt N}$)

\section{p-spin model}
\renewcommand{\theequation}{\thesection.\arabic{equation}}
\label{sect4}

The spherical $p-spin$ model \cite{CrsoI}, unlike the previous two, has
many (exponentially with $N$) \cite{CrsoII} minima.  
The Parisi ansatz for the replica  solution has a one-step replica 
symmetry breaking.

The long time out of equilibrium dynamics \cite{Cuku} has, again, 
in the low-temperature phase, two regimes of times:

i) for $t,t' \; \rightarrow \infty$ and $t-t'$ finite the correlation
and response functions are time-translational invariant, and satisfy
FDT.  At {\em zero temperature} and  large $t-t'$, but still $<<t$,
also for this model:
\beq
C_{FDT} \; \simeq \;  q_{EA}+ A (t-t')^{-1/2} \;\;\; ; \;\;\; R_{FDT} 
\propto (t-t')^{-3/2}
\label{4.1}
\eeq
The relaxation exponents change with temperature and are given in
\cite{Crhoso}.

The dynamical Edwards Anderson parameter ($T \sim 0$)
\beq
\frac{1-q_{EA}}{T}= {\sqrt \frac{2}{p(p-1)}} 
\label{edan}
\eeq
as well as the asymptotic `threshold' energy density
\beq
e_{thres} =-{\sqrt \frac{2(p-1)}{p}}
\label{thres}
\eeq
are  {\em different} from their corresponding Gibbs-measure counterparts. 
We shall re-derive them from a geometrical point of view below.

ii) For widely separated times, the correlation becomes smaller than
$q_{EA}$, and  is not time-translational invariant.  The response
function in this regime also violates FDT, and, unlike the previous
case, yields a long-time memory with aging effects that are similar to
those experimentally observed in real spin-glasses.

{\bf Structure of minima}

The system has many minima, ranging in energy density from the ground state 
$e_{Gibbs}$ (as calculated in the replica calculation) up to a  
threshold energy $e_{thres}$
\cite{Kupavi,Cuku,CrsoII} (\ref{thres}). 
The corresponding Edwards-Anderson parameters of these states are given, 
in terms of the energy density of each minimum $e$ by
\cite{Kupavi}:
\beq
\frac{1-q_{EA}}{T}= \frac{1}{p-1} \left\{ -e - (e^2-e_{thres}^2)^{1/2} \right\}
\label{titi}
\eeq

The asymptotic energy, as well as the dynamical Edwards-Anderson parameter 
tend to the values (\ref{thres}),(\ref{edan}) corresponding to the threshold 
states, though {\em the system never relaxes into any of these} \cite{Cuku}.

Using the equation of motion, and the spherical constraint, one has that 
the Lagrange multiplier $z(t)$ is related to the energy density by:
\beq
z(t)= - p \;  e(t)
\label{4.5}
\eeq

The energy Hessian can be calculated by expanding up to second
order around a stationary point and using the spherical constraint,
to get (cfr. (\ref{mmf})):
\beq
H_{ij}=
\frac{\partial^2 E}{\partial s_i \partial s_j} + z(t) \delta_{ij} 
\label{4.6}
\eeq

Using the homogeneity in the expression for the energy, we first note that 
the equation of stationarity  implies that ${\bf H}$ has an
eigenvector in the `radial' direction $s_i$ of eigenvalue $-p(p-1)z(t)$. 

We can find the spectrum in the directions orthogonal to this one by using 
the `locator expansion' \cite{Brmo}, or directly by noting that 
$\frac{\partial^2 E}{\partial s_i \partial s_j} $ is a sum of many 
($ \sim N^{p-2}$) terms, and hence assuming that in the directions orthogonal 
to $\vec{s}$ the couplings can be taken as uncorrelated from the configurations
in the large $N$ limit (this is indeed the assumption in \cite{Brmo}).  
The mean-squared element can be obtained from:
\beqa
\overline { [\frac{\partial^2 E}{\partial s_i \partial s_j} ]^2} &=& 
\left [\frac{p(p-1)}{p!} \right]^2
\sum_{i_1,...,i_{p-2} \neq i,j} \sum_{j_1,...,j_{p-2} \neq i,j} 
\overline{ J_{i_1,...,i_{p-2}} J_{j_1,...,j_{p-2}} }
s_{i_1} ... s_{i_{p-2}} s_{j_1}...s_{j_{p-2}} \nn
&=& \left [\frac{p(p-1)}{p!} \right]^2 N^{p-2} 
\left( \frac{p!}{2N^{p-1}} \right) (p-2)! \nonumber \\
&=& \frac{p(p-1)}{2N}
\label{4.8}
\eeqa
where we have used the variance of the couplings, and the factor $(p-2)!$ 
counts the number of ways of matching the $\{i_k \}$ with the $\{ j_k \} $.
The matrix of second derivatives is then a random matrix whose distribution of
eigenvalues is a semicircle law with support in 
$[ -\sqrt{2p(p-1)},+\sqrt{2p(p-1)}]$, plus a projector in the 
direction of $\vec{s}$.

Using  (\ref{4.5}), (\ref{4.6}), (\ref{4.8})) and (\ref{thres})
we find that  the  spectrum of the Hessian $\rho_e(\lambda)$ in the direction 
tangential to the constraint is a shifted semicircle law with support in
\beq
p(e_{thres}-e) < \lambda < -p(e_{thres}+e) \;\;\; ; \;\;\; 
\lambda_{min}=p(e_{thres}-e)
\label{spectrum}
\eeq

A direct calculation using:
\beq
\frac{1-q_{EA}}{T} = \int d \lambda \frac{\rho_e(\lambda)}{\lambda}
\eeq
yields back (\ref{titi}). A more complete  confirmation of (\ref{spectrum}) 
can be  obtained by making a low-temperature expansion around a minimum and 
checking that using the form of $\rho_e(\lambda)$ one
gets the same results as in the TAP approach of \cite{Kupavi}.

We now understand the origin of the threshold level: the `gap' in the spectrum 
of the Hessian becomes smaller as one considers states that are higher, 
until it disappears at the $e=e_{thres}$.
Above the threshold the gap is negative: those critical  points are unstable 
and their index increases with increasing energy.

Finally, we can estimate the energy difference between the critical points of 
index $I$ and the highest threshold minima.
Using $\lambda_{min}$ from (\ref{spectrum}) and the semicircle law 
one easily gets:
\beqa
I = N \int^0_{\lambda_{min}} d\lambda \  \rho_I ( \lambda ) 
&\simeq & N \int_0^{p (e_I - e_{thres})} d\lambda \  \lambda^{1/2} \nonumber \\
I \propto N (e_I - e_{thres})^{3/2} \ \ \
& \Rightarrow & \ \ \ E_{I}-E_{thres} \propto I^{2/3} N^{1/3}
\eeqa
In particular, we have for the barriers separating threshold minima ($I=1$):
\beq
E_{I=1}-E_{thres} \sim N^{1/3}
\eeq

{\bf Out-of equilibrium Dynamics}

In order to study the  long time dynamics starting from a random configuration,
we  start by considering the Hessian given by (\ref{ccc}) and (\ref{mmf}).  
The calculation of  the {\em time-dependent} spectrum of the second derivatives 
of the  energy is obtained by making as in the preceding section
the assumption of independence of the configurations and couplings in the 
large-$N$ limit. Repeating the calculation (\ref{4.6}) for this case, 
we find that the Hessian consists of a  random matrix of elements with variance 
as in (\ref{4.8})  plus a shift term.  The eigenvalue density $\rho_t(\lambda)$ 
is then a semicircle law with support in:
\beq
p(e_{thres}-e(t)) < \lambda < -p(e_{thres}+e(t)) \;\;\; ; \;\;\; 
\lambda_{min}(t)=p(e_{thres}-e(t))
\label{spectrum2}
\eeq
This assumption is confirmed numerically in Fig.5 , where we plot the 
integrated spectrum of the Hessian for different times. 
In the inset we show the integrated spectrum of the matrix of second derivatives 
of the energy (i.e. the spectrum of {\bf H} minus the shift) and compare it with
an integrated semicircle law.

\begin{center}
   \parbox{8.5cm}{\epsfig{file=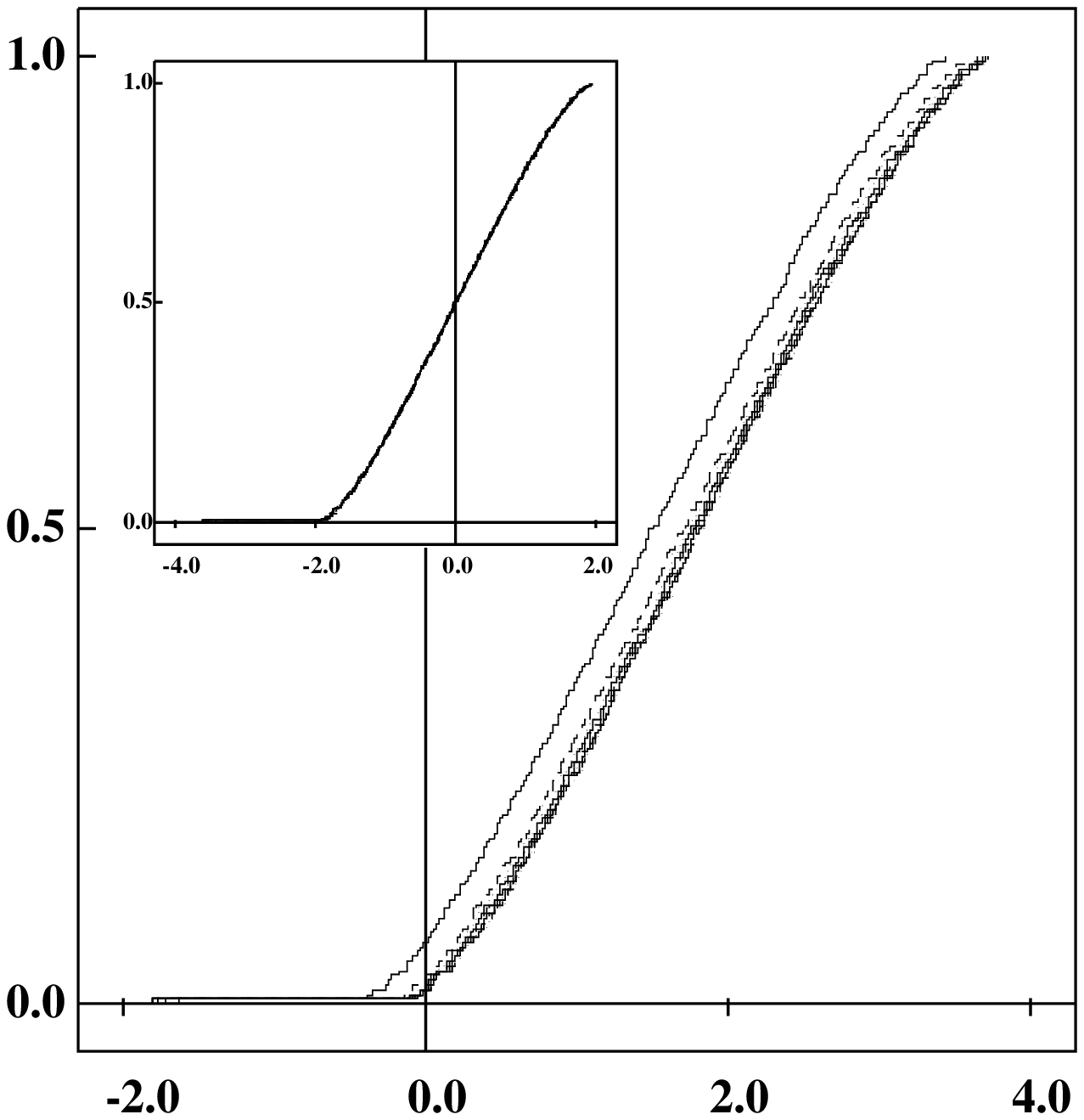,width=8cm}}\\
FIG. 5.	Integrated eigenvalue ($\lambda / \sqrt 3$) distribution of the 
	Hessian at 10 different times for p=3 and N=200.
     Inset: the shifted distribution for all times and the analytical
	    integrated semi-circle law.
\end{center}

We are now in a position to rederive some results for the out of equilibrium 
dynamics of \cite{Cuku}: If we now claim, as in the preceding sections, that 
because of the dominance of borders the dynamics is such that the Hessian 
has a (decreasing) number of negative eigenvalues at all finite times, 
we reobtain the `marginality condition':
\beq
\lim_{t \rightarrow \infty} e(t) = e_{thres}
\eeq

At this point it is important to remark that this last equation does 
{\em not} mean that the system relaxes  into a near-threshold state: at all 
finite times an infinite system has a Hessian with an {\em infinite} number 
of directions in which the energy is a maximum.
If at a given finite time the system is close to a border $\partial_I$, 
we may ask how many different basins meet there, i.e. to what extent the 
system is `almost undecided'. A study \cite{tommasso} of the 
random partitioning of a high-dimensional space suggests that generically
$I+1$ basins meet at $\partial_I$ - an infinite quantity at finite times.

We have seen that the saddles separating threshold minima are typically 
$O(N^{1/3})$ above the threshold level, while the energy is at all finite 
times $O(N)$ above this level. Again, we confirm that at all finite times 
the constant-energy surface is {\em not} disconnected into components.

Up to now the system seems to behave dynamically in a very similar fashion 
to the $p=2$ model.  An important difference appears when we consider the 
evolution of the eigenbasis of the Hessian.
Because the actual elements of the second derivative matrix now  depend on 
the time via the time-dependence of the spins, we may expect that unlike the 
case $p=2$ the eigenbasis of $H$ also depends on time.
Indeed, (see Fig.6) the overlap of the eigenvectors at two different times 
($t,t'$ such that $C(t,t') \sim 0.7$) are very small.
Although the spectrum of the Hessian leads us to an image of a `channel' 
whose characteristics change slowly with time (as in the case $p=2$), we now 
see that such a channel twists and turns chaotically with time.

\begin{center}
   \parbox{9cm}{\epsfig{file=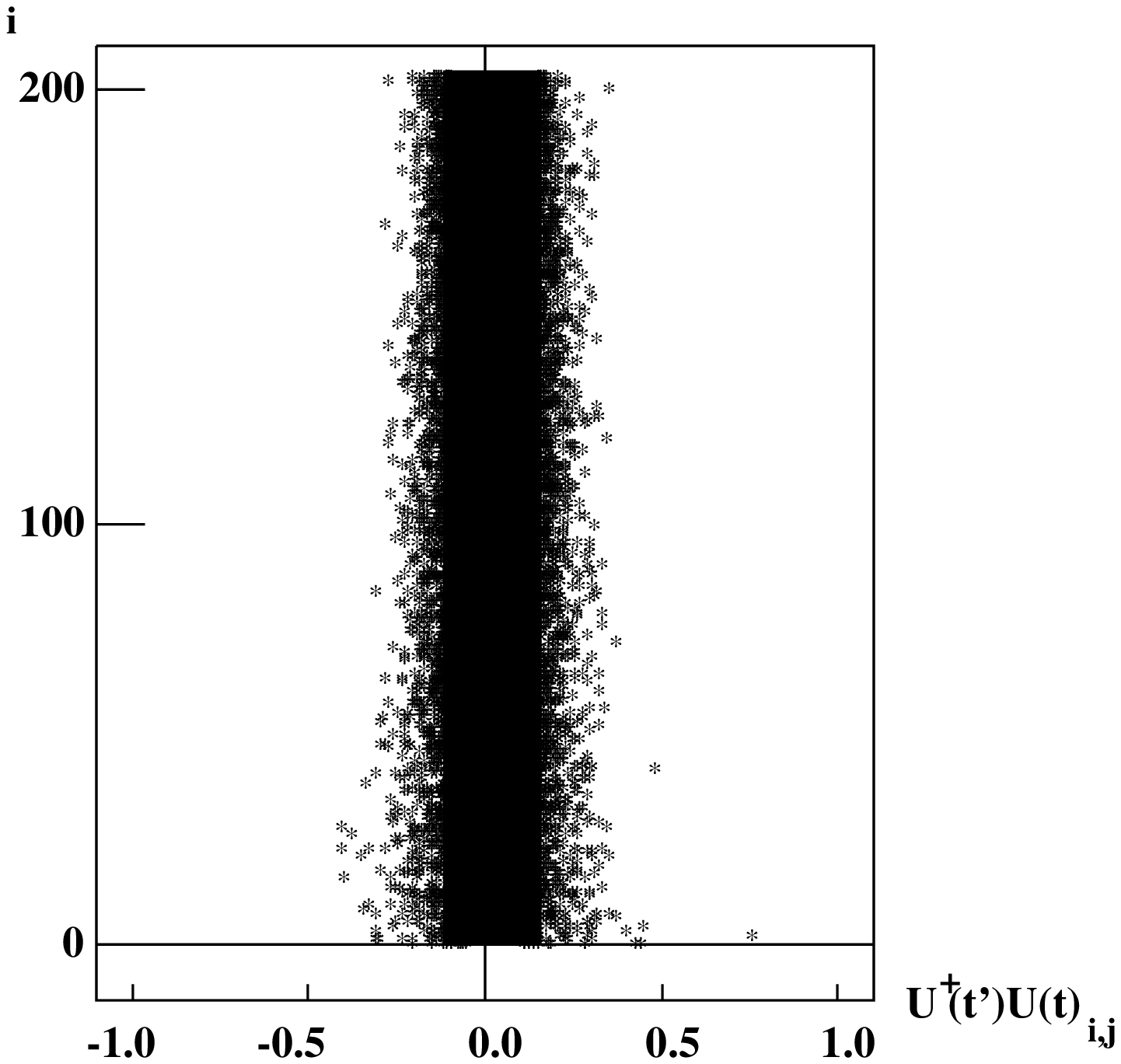,width=8.5cm}}\\
FIG. 6.	The statistics of overlaps between the eigenvectors of
	the Hessian, at two times 
	($[U^{\dagger}(t')U(t)]_{i,j} \ vs \ i$), 
	for p=3 N=200.
\end{center}

Let us now turn to the analysis of the `fast' relaxations
(the quasi-equilibrium regime). We have seen in the preceding section that what 
is relevant for this regime is the spectrum of the time-averaged Hessian 
(cfr. (\ref{3.18}) (\ref{3.19}) and (\ref{corr2})).
Since for this problem the eigenbasis of the Hessian turns, in principle we have
to take care of the fact that the spectrum of the time-averaged Hessian is not 
the same as the time average of the spectrum. 

The spectrum of the averaged Hessian can still be obtained with the same 
assumptions, but taking into account in (\ref{4.8}) that the spins are evaluated
at two different times.  This will introduce a factor $\sim C(t,t')^{p-2}$.
However, since we are here  interested in the regime of times for which 
$C(t,t')>q_{EA}$, and we are restricting ourselves to 
{\em near-zero temperatures}, the effect of the variations of the configuration 
(and hence the `turning' of the Hessian) can be neglected.
Following then a calculation as in the preceding section, 
we reobtain (\ref{4.1}).

Hence, we see that the fact that the form of the steepest `walls' of 
the `channel' preserve their form (even if the channel twists) explains 
the validity of the fluctuation-dissipation theorem and time-translational 
invariance in the quasi-equilibrium regime.

\section{Conclusions}
\label{sect5}

In this paper we have argued that the relaxational dynamics of a system in
the thermodynamic limit has characteristics that  lead naturally to slowing
down and  are  a direct consequence of 
the infinite-dimensionality of phase-space.

 The peculiarities of infinite-dimensional geometry may be overlooked 
if one seeks for inspiration from a low dimensional phase-space sketch: 
such a sketch for a ferromagnet would consist of a double well and would 
lead us to conclude that after a rapid quench the ferromagnet `falls' into 
one of the states, though we know that this does not happen in finite times.

In the case of a ferromagnet, we are kept from jumping into wrong conclusions
by a  picture of snapshots of the domain structure at different times - 
a naturally infinite-dimensional description of a configuration.
At present it is not known if every form of slow dynamics present in nature is 
just some sophisticated version of domain growth 
(cfr. the long-lasting controversy with spin-glasses):
hence the interest of trying to explore the consequences of 
infinite-dimensionality of phase-space directly, without invoking 
real-space structures. 

In order to isolate these phase-space geometric causes of slowing-down 
(which do not involve rapid jumps) from barrier-crossing mechanisms, 
we have deliberately concentrated on systems which have a non-Arrhenius 
behaviour at near-zero temperatures.
On the other extreme, there is a  picture by Bouchaud \cite{Bo} of 
jumps between phase-space traps which is at the same time simple and 
yields excellent results for spin-glasses.

 Whichever turns out to be the complete description of the problem of slow 
dynamics, it  will  have to take into account all elements, blending
continuous and  discontinuous trapping together in a single picture - 
possibly with each correlation scale dominated by one type of mechanism. 

Let us finally remark that though it seems important to have a 
phase-space intuition of what happens with a  system that ages, 
it will not substitute a dynamical computation:
by the time one introduces all the necessary elements that go into the 
definition of a `barrier', one has taken into account all the  paths leading 
to it and their respective probabilities: precisely what one computes in a 
dynamical calculation.
Moreover, if as we have argued here the saddle-points are also relevant, 
in order to know their structure we are faced with a Morse-theory problem 
that is again best studied \cite{Wi} using methods that are closely related
to Langevin dynamics.

\acknowledgments

We wish to thank J.P. Bouchaud, L.F. Cugliandolo and M. Mezard for discussions 
and suggestions.


\begin{thebibliography}{99}
%
\bi{St} L. C. E. Struik; `{\it Physical aging in amorphous polymers and
other materials}', Elsevier, Houston (1978).
%
\bi{Aging} L. Lundgren, P. Svedlindh, P. Nordblad and O. Beckman;
{\sl Phys. Rev. Lett.} {\bf 51} 911 (1983). \newline
M. Alba, J. Hammann, M. Ocio and Ph. Refregier; {\sl J. Appl. Phys.}
{\bf 61} 3683 (1987).
%
\bi{Fihu} D. S. Fisher, D. Huse; {\sl Phys. Rev. B} {\bf 38}, 386 (1988); \n
D. S. Fisher and D. Huse; {\sl Phys. Rev. B} {\bf 38}, 373 (1988); \n
G. J. Koper and H. J. Hilhorst; {\sl J. Phys. France} {\bf 49}, 429 (1988).
%
\bi{Bo} J-P Bouchaud; {\sl J. Physique} {\bf 2}, 1705 (1992). \n
 J.P. Bouchaud, D.S. Dean; {\sl J. Physique I (France)} {\bf 5}
(1995) 265.
%
\bi{Siho}P. Siban; {\sl Phys. Rev. B} {\bf 35}, 8572 (1987);\n
K. H. Hoffmann and P. Sibani; {\sl Z. Phys. B} {\bf 80}, 429 (1990).
%
\bi{Nest}  C.M. Newman, D.L. Stein;  {\sl Phys. Rev. E} {\bf 51}, 5228 (1995).
%
\bi{Babume} A. Barrat, R. Burioni and M. Mezard; {\bf cond-mat 9509142},
submitted to {\sl Journal of Physics A}.
%
\bibitem{Bray} A.J. Bray, {\sl Adv. in Phys.} {\bf 43}, 357 (1994)
%
\bi{Kothjo}
J. M. Kosterlitz, D. J. Thouless and R. C. Jones; {\sl Phys. Rev. Lett.}
{\bf 36}, 1217 (1976).
%
\bi{Cidi} S. Ciuchi and F. de Pasquale; {\sl Nucl. Phys. B} {\bf 300} [FS22], 
31 (1988).
%
\bi{Cude}  L.F. Cugliandolo, D.S. Dean; {\sl J. Phys. A} {\bf 28}, 4213 (1995).
%
\bi{CrsoI} A. Crisanti, H-J Sommers; {\sl Z. Phys. B} {\bf 87}, 341 (1992).
%
\bi{Cuku} L.F. Cugliandolo and J. Kurchan; {\sl Phys. Rev. Lett.} {\bf 71}, 173
(1993)
%
\bi{Kupavi} J. Kurchan, G. Parisi and M. Virasoro;
  {\sl J. Physique I (France)} {\bf 3} (1993) 1819.
%
\bi{CrsoII}  A. Crisanti, H-J Sommers; {\bf cond-mat 9406051}.
%
\bi{Derrida} The fact that initial random conditions are near borders was 
already noted in the study of fracture; see 
B. Derrida; {\sl J. Phys A} {\bf 20} (1987) L721.
%
\bi{Pa} Some consequences of this are discussed in: 
G. Parisi; {\bf cond-mat 9412034}.
%
\bi{Hofr} H. Horner; {\sl Z. Phys. B} {\bf 66}, 175 (1987). \n
M. Freixa-Pascual and H. Horner; {\sl Z. Phys. B} {\bf 80}, 95 (1990).
%
\bi{martorell} There is a large bibliography on quantum wires, see e.g.: \n
D.W.L. Sprung, Hua Wu and J. Martorell; {\sl J. Appl. Phys.} {\bf 71} (1),
515 (1992).
%
\bi{Me} M. L. Mehta; {\it `Random matrices and the statistical theory of 
energy levels'} \n (Academic, New York, 1967).
%
\bi{Crhoso}
 A. Crisanti, H. Horner and H.-J. Sommers; {\sl Z. Phys. B}
{\bf 92}, 257 (1993). 
%
\bi{Brmo} A. Bray and M.A. Moore; {\sl J. Phys. C} {\bf 12}, L441 (1979).
%
\bi{tommasso}  T. Aste and N. Rivier; {\sl J. Phys. A} {\bf 28}  (1995) 1381.
%
\bibitem{Wi} E.Witten; {\sl Nucl. Phys. B} {\bf 202}, 253 (1982).
%
\end{thebibliography}
\end{document}